\documentstyle[preprint,aps]{revtex}
\begin{document}
\draft

\title{Power-Law Behaviors in Nonlinearly Coupled Granular Chain under Gravity}
\author{Jongbae Hong and Heekyong Kim}
\address{Department of Physics Education, Seoul National University, Seoul 151-742, Korea}
\maketitle
\begin{abstract}
We find power-law  behaviors of grain  velocity in both  propagation and  backscattering
in a gravitationally compacted granular chain with nonlinear contact force. We focus on the
leading peak of the velocity signal which decreases in a power-law $d^{-\alpha}$,
where $d$ is the location of the peak, as the signal goes
down.
The ratio of backscattered to incident leading velocity also follows
a power-law $d_i^{-\beta}$, where $d_i$ is the depth of impurity.
The up-going backscattered signal is nearly solitary. Therefore, the overall change of
the leading velocity peak is given by the power-law $d_i^{-(\alpha+\beta)}$.
We find $\alpha=0.250$ and $\beta=0.167$ for the Hertzian
contact force.

\end{abstract}



Recently, physics   of granular  materials  attracts great   interest, since  the  materials are 
ubiquitous  around us and their properties are unique and also useful in many applications\cite{nagel0}. 
The propagation of  sound or  elastic wave  in granular  medium is also  one of  interesting 
subjects related to 
the properties of  granular matter[2].  A rather  simple system,  the granular  chain with 
nonlinear contact 
force including   Hertzian contact\cite{landau},  has  been revived   by finding  a  soliton in 
transmitting elastic impulse. This soliton
found in a loaded horizontal  Hertzian chain was first predicted  by Nesterenko\cite{nest} and 
its experimental verification was  performed by Lazaridi  and Nesterenko\cite{lazar} and  recently 
by Coste {\it et al}.\cite{coste}. Even though three-dimensional granular systems may not follow 
simple Hertzian contact force law due to geometrical effect\cite{goddard}, the simple nonlinear 
contact law is still interesting because the existence of solitary wave may provide a possibility to get 
information inside granular matter and the one-dimensional system usually provides the starting point for 
studying more complex systems.

In addition to studying  horizontal chain, the  gravitationally compacted chain  and column of 
grains have been studied  by  Sinkovits  and  Sen\cite{sinko1,sinko2,sen}   in terms   of molecular   
dynamics simulations. 
The two cases are not the same in the following sense that the former shows robust solitary 
wave within certain amount of loading, while the latter shows solitonlike wave whose waveform  disperses 
little by little 
as it propagates down. Sen {\it at el}.\cite{sen} studied backscattering of the solitonlike wave 
by impurity in a   gravitationally compacted   chain to  get  the  information on   buried impurity   such 
metal-poor land mines. They obtained interesting relations between the  phase of grain-velocity signal and the  
mass of impurity. 
The result  reflects  the well-known  phase-density  relation in   the connected string   with 
different densities. 

In this Letter, we focus on  the propagating and backscattering characteristics  of the leading 
part of 
grain-velocity signal in  the gravitationally  compacted granular chain  with Hertzian  contact 
force especially. 
We also treat other types of nonlinear forces. We report 
interesting power-law behaviors  of the  velocity signal. This  work is  attractive because it 
supplies new  
laws of signal  propagation in  the gravitationally  compacted granular  chain with  nonlinear 
contact force and 
a possibility of identifying the location  of buried impurity by observing  the peak-velocity of 
the leading part of 
returned signal at surface. 

To study the dynamics of grains in the gravitationally compacted Hertzian chain,  
we solve numerically the equation of motion of a grain at $z_i$,
\begin{eqnarray}
m\ddot{z}_i&=&\frac{5}{2}a[\{\Delta_0-(z_i-z_{i-1})\}^{3/2}-\{\Delta_0-(z_{i+1}-z_{i})\}^{3/2}]
\nonumber \\ 
&-&mg,
\end{eqnarray}
where $m$ is the mass of grain, $\Delta_0=R_i+R_{i+1}$, and $a$ is  the constant defined in 
Eq. (2) below. We 
neglect plastic deformation in Eq. (1). This equation 
of motion comes from the  Hertzian interaction energy between neighboring  granular spheres 
which is 
given by\cite{landau}
\begin{equation}
V(\delta_{i,i+1})=\frac{2}{5D}\left(\frac{R_iR_{i+1}}{R_i+R_{i+1}}\right)^{1/2}\delta^{5/2}_{i,i+1}
\equiv a\delta^{5/2}_{i,i+1},
\end{equation}
where $\delta_{i,i+1}$ is the overlap between two adjacent spherical grains and is  given 
by $\delta_{i,i+1}=R_i+R_{i+1}-r_{i,i+1}$, in which  $R_i$ is the  radius of the  sphere whose 
center is at position $i$ and $r_{i,i+1}$ is the distance between $i$ and $i+1$, and
\begin{equation}
D=\frac{3}{4}\left(\frac{1-\sigma^2_i}{E_i}+\frac{1-\sigma_{i+1}^2}{E_{i+1}}\right),
\end{equation}
where $\sigma_i$, $\sigma_{i+1}$ and  $E_i$, $E_{i+1}$  are  Poisson's ratios and Young's 
moduli 
of the bodies  at neighboring  positions, respectively\cite{landau}. The equilibrium condition
\begin{equation}
g\sum_{j=i+1}^N m_j=\frac{5a}{2} \delta^{3/2}_{i,i+1}
\end{equation}                                      
has been used for the $i$-th grain of a chain of $N$ grains. 
A criterion for the initial impulse exists to make Eqs. (1) and (2) valid\cite{coste,johnson}. 
Therefore, in this work, we choose comparatively weak impulses to satisfy this criterion. 

Our system is a vertical chain of $N=10^3$ grains which has five consecutive impurity grains 
which are 
different only in mass from other spherical grains. The  mass of impurity is taken as half of 
the mass of medium grains. 
As a calculational tool, we use the third-order Gear predictor-corrector algorithm\cite{gear}
and the same program units as those used in Ref. \cite{sen}. That is, the units of distance,
mass, and   time are   $10^{-5}$m, $2.36\times   10^{-5}$kg, and   $1.0102\times 10^{-3}$s, 
respectively.
These units gives the gravitational acceleration $g=1$.  We set the grain diameter 100, mass 
1, and the constant $a$ of Eq. (1) 5657 for molecular dynamics simulation.

We focus on the peak value  of the very leading velocity signals  from the moment of initial 
impulse imposed on the first grain at top of chain to the moment of return to the first grain.
Figure 1(a) shows  the snap  shots of  propagating velocity signal  passing different  depths. 
Initial impulse velocity $0.1$ in program units  has been applied to  the first grain. One  
can easily see from Fig. 1(a) that more grains are involved in the signal, i.e., signal disperses 
as time passes. We find that even though loading on the grain increases in depth due to gravity, 
total kinetic energy and potential energy of the grains involved in the signal remains constant
independently. Therefore, the decrease of peak  velocity results from only dispersion of the signal. 
Fig. 1(a) shows that  the  propagating  signal  is  not solitary   in the gravitationally 
compacted   chain.  This   is   quite   different  from   the   case   of  loaded   horizontal 
chain\cite{nest,coste}. 

In this work,  we are  interested in  the effect  of gravity  which changes  loading of chain 
continuously and 
the time- or depth-dependent behavior of the leading peak shown in Fig. 1(a). 
Fig. 1(b) is the log$_{10}$-log$_{10}$ plot of the leading velocity peak versus its location. 
A clear power-law behavior is seen in Fig. 1(b). 
The explicit expression for the leading peak velocity is given by 
\begin{equation}
\frac{v(d)}{v_i}=Ad^{-\alpha},
\label{propa}
\end{equation}  
where $v_i$ is the initial velocity, $d$ is the distance from top to the leading peak, 
and $A$ is the intersection which is not equal  to one because of boundary effect at the  top 
of chain.
But the signal quickly goes to the stable form shown  in Fig. 1(a). We will show this  
in Fig. 2 . 
Figure 1(b) obtained for Hertzian contact gives rise to the power-law index $\alpha=0.250$.
Figure 2 drawn in log$_{10}$-log$_{10}$ plot shows the propagating behaviors of the leading 
peak of the velocity signal for various 
nonlinear contact forces, such as $n=5/2, 3, 7/2,$ and $4$ of the potential $V(\delta_{i,i+1})=
a\delta^{n}_{i,i+1}$. The abscissa of Fig. 2 denotes elapsed time after impulse. 
One can see that the initial impulse quickly follows the power-law of Eq. (\ref{propa}) in 0.1 
second which corresponds to passing  40th grain in length.  Interesting things in  Fig. 2 are (1)  the 
slopes are very close to each other even though there is crossing, therefore, one expects similar 
propagating  behaviors  for  these   nonlinear contact   forces,  and  (2)  Hertzian   case is 
well-separated from other nonlinear cases. It is unclear right now why Hertzian is special.

The signal  performs  backscattering when   it meets impurity.   Figure 3(a)  shows typical 
backscattering 
by five consecutive impurity grains  of mass 0.5 and  depth 500 in program  units. The total 
energy of 
the incident signal is divided into transmitted and reflected parts, i.e., $v_{inc}^2\approx
v_{trans}^2+v_{back}^2$. The backscattered leading  peak diminishes abruptly  compared with 
second leading peak in the process of scattering. We call the ratio of backscattered to incident 
leading peak the reflection coefficient of the leading peak velocity for present analysis.
We find a remarkable phenomenon that the  reflection coefficient of the leading peak velocity 
also follows a power-law in depth of impurity. The reflection coefficient, of course, depends on the 
mass of impurity. We  restrict this  study to  the case  of impurity mass  0.5 while  the mass  of 
medium grain is 1.
Fig. 3(b)  is the  log$_{10}$-log$_{10}$ plot  of the  reflection  coefficient, i.e.,  the ratio  of 
backscattered 
leading peak to incident one  versus the location of  impurity $d_i$. A clear power-law  is seen in 
Fig. 3(b) and the expression is given by 
\begin{equation}
\frac{v_b(d_i)}{v(d_i)}=Bd_i^{-\beta},
\label{back}
\end{equation}
where $v_b(d_i)$ is  the leading  peak velocity  after backscattering by  impurities at  depth 
$d_i$ and $B$ is the intersection of log$_{10}$-log$_{10}$ plot. 
Figure 3(b) obtained for Hertzian contact gives rise to  the power-law index $\beta=0.167$ for 
the impurity mass 0.5. We  take the value  of $v_b(d_i)$  at the moment  of separation from  the tail  
of incident signal. We confirm that conservation laws of energy and momentum are always retained. 

This power-law  behavior is  an unexpected  result. We  thought this  may stem   from the 
change of loading as 
the location of impurity changes in the gravitationally compacted chain. We checked this idea 
for the 
horizontal chain by changing the amount of loading. We found that there is no change in the 
reflection 
coefficient upon loading for the  solitary wave in the  horizontal chain. We also  confirm that 
the amplitude of 
incident solitary  wave in  horizontal  chain cannot  change the  reflection  coefficient either. 
Therefore, 
we conclude   that the   power-law behavior   in Eq.   (\ref{back}) is   a characteristic   of 
nonsolitariness of the wave in gravitationally compacted granular chain.

We also check the propagation of backscattered wave. It is interesting to see that 
the backscattered waveform is very solid and it shows negligible change in the height of the 
leading peak within the depth we considered in this work. Also interesting thing is that the 
propagating behavior of the backscattered velocity signal is not power-law but linear in depth 
as shown in Fig. 4 drawn in absolute scale. The saw tooth form stems from finite radius of grain.
The trend of changing leading peak in Fig. 4 is flat in the velocity scale used in this analysis. 
We find that the number of grains participating in the signal increases in the case of signal 
going down, while it remains nearly constant in the up-going situation. Therefore, there is little
dispersion when the signal goes up in the gravitationally compacted chain. 
In addition, no kinetic and potential energy change either in this case.  
The linear-law of Fig. 4 does not play a role for the shallow region but play an appreciable 
role when backscattering produces at a deep enough place.

As a summary  of the  above analysis,  we obtain  the overall  power-law relation  between 
returned velocity $v_r$ at top and the depth  of impurity $d_i$. Since the kinetic and contact 
potential energy is transferred to the kinetic and gravitational potential energy of the 
first grain at top, the leading backscattered peak velocity at grain 2, i.e., 
$v_b(2)\approx -0.004$ in Fig. 3(a), is quite different from the leading peak velocity at grain 1
which we define the returned leading peak velocity $v_r$. Figure  5 shows the log$_{10}$-log$_{10}$  
plots of $v_r/v_i$ versus the location  of  impurity $d_i$  for  different contact  forces  and 
initial  impulse  velocities $v_i=0.1, 0.05,$ and $0.01$ in program units. One can see a big 
difference between $v_r$ and $v_b(2)$. We separate artificially the lines of Fig. 5  into two groups by 
dividing the data of $V(\delta_{i,i+1})=a\delta^{3}_{i,i+1}$ by  1.1 to avoid overlap  with those of  
Hertzian interaction $V(\delta_{i,i+1})=a\delta^{5/2}_{i,i+1}$. 
One  can see  that the  power-law is  independent of initial impulse. 
The equation of the straight lines of Fig. 5 is written as
\begin{equation}
\frac{v_r}{v_i}=R(v_i) d_i^{-(\alpha+\beta)},
\label{law}
\end{equation}                               
where the coefficient $R(v_i)$, the intersection of the log$_{10}$-log$_{10}$ plot, contains the 
energy transfer at top and the coefficients $A$ and $B$ of Eqs. (5) and (6), respectively.
$R(v_i)$'s weakly depends on the amount of initial impulse and the type of contact force law as shown 
in Fig. 5. The power-law index $\alpha+\beta=0.417$  for Hertzian contact  and $0.378$ 
for another. From the analysis above, we know that the index is composed of 
two different sources, i.e.,  the index for  downward propagation $\alpha$  and the index for 
backscattering $\beta$. 
One can confirm this by adding the slopes of Fig. 1(b) and Fig. 2(b) for Hertzian chain. 

In conclusion, we solve numerically the many-body equations of motion of the  gravitationally 
compacted granular chain coupled by nonlinear contact  forces and analyze the propagating  
behaviors of the leading peak 
of the velocity signal and the backscattering characteristic by buried impurity. We find two
power-law relations in Eqs. (\ref{propa}) and (\ref{back}) for propagation and backscattering, 
respectively. The former contains only the effect of dispersion of nonsolitary wave as it 
propagates. The latter, however, has nothing to do with dispersion but the phenomenon 
occurring at the point of scattering. Comparing with the case of solitary wave 
in loaded horizontal chain, we conclude that Eq. (\ref{back}) is a special feature of nonsolitary 
property of the gravitationally compacted granular chain. 
The backscattered up-going wave, on the other hand, is a nearly solitary wave and 
does not follow power-law but follow very weak linear-law unlike the down-going wave.  We conclude 
that this interesting feature is the characteristic of the gravitationally compacted granular chain
with nonlinear contact force. The solitary behavior of the  up-going 
wave in the gravitationally compacted granular chain may be 
useful to find buried impurity like nonmetallic land mine. 

In Eq. (\ref{law}), we obtain the combined power-law on the location of impurity 
for the returned leading peak velocity  at top of chain. This power-law  also exists for other 
nonlinear contact forces as well as Hertzian as we show  in Fig. 5. We therefore conclude that above  
features and power-laws are generic   for   any   nonlinear   contact   forces   of   type   
$F(\delta_{i,i+1})\propto \delta^{n-1}_{i,i+1}$ in the gravitationally compacted granular chain.

One of authors (J.H.) thanks Prof. S. Sen of State University of New York at Buffalo for 
useful discussions on this work. This work has been supported by KOSEF and partially by 
BSRI 97-2420 of the Ministry of Education.

\newpage
{\bf Figure Captions}

\begin{description}
\item[Fig. 1] :  (a) Snap  shots of  propagating wave  forms in  a gravitationally  compacted
granular chain with Hertzian contact  law. (b)  Log$_{10}$-log$_{10}$ plot  of $v(d)$  vs. depth $d$
for the leading peak in (a). Slope of the straight line is $-0.250$.
\item[Fig. 2] :  Log$_{10}$-log$_{10}$ plot of  $v(d)$ vs.  $d$ for the  leading peak for
various interaction potentials, i.e., $n=5/2, 3, 7/2,$ and $4$ of $V(\delta_{i,i+1})=a\delta^{n}_{i,i+1}$.
\item[Fig. 3] : (a) Snap shot of scattering by five light
impurities at 500.
(b) Log$_{10}$-log$_{10}$ plot of  $v_b(d_i)/v(d_i)$ vs. $d_i$.  Slope of the straight  line is
$-0.167$.
\item[Fig. 4] : Plot of leading peak vs. its location for
the backscattered part of  the velocity
signal. Absolute scales are used for both abscissa and ordinate. Five impurities are at 700.
\item[Fig. 5] :  Log$_{10}$-log$_{10}$ plot of  $v_r/v_i$
 vs.  $d_i$ for the leading peak
for various initial velocities, $v_i=0.1, 0.05, 0.01$. Upper three lines are Hertzian ($n=5/2$) cases.
Lower three lines ($n=3$) are lowered artificially. Slopes are 0.416 ($n=5/2$) and 0.378
($n=3$).

\end{description}

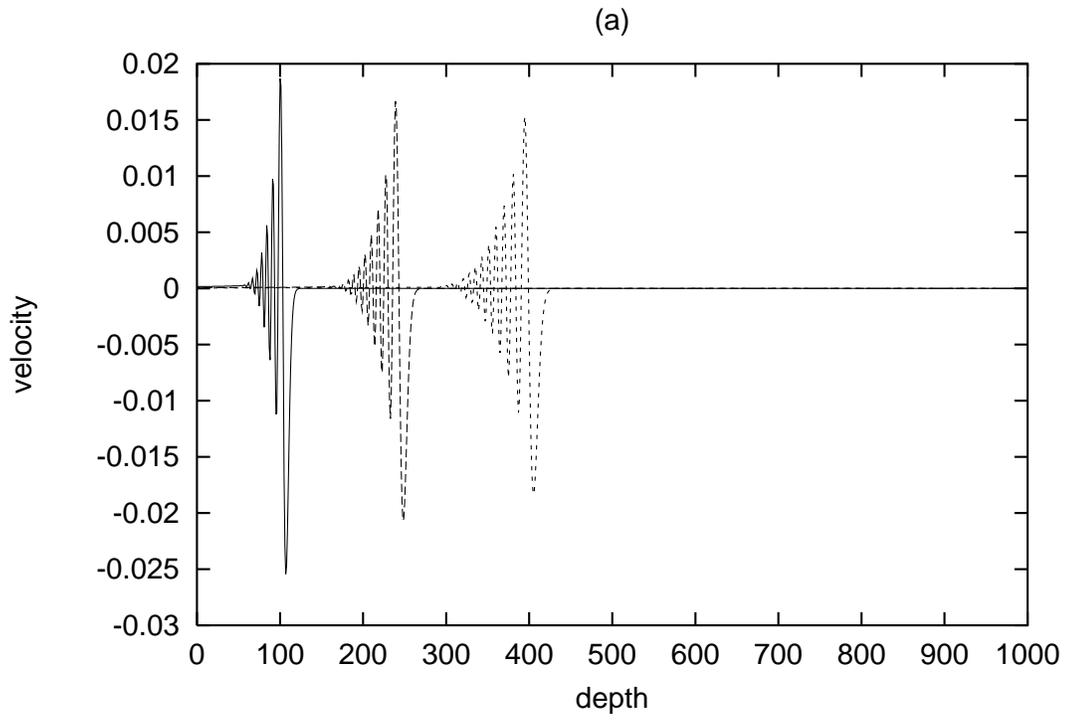
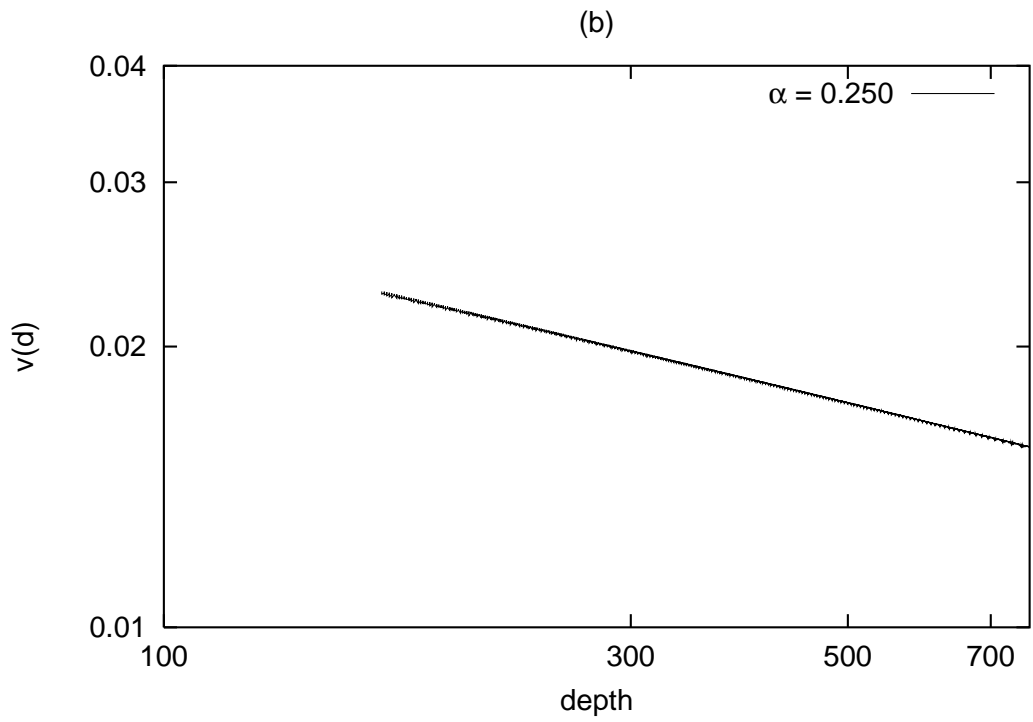
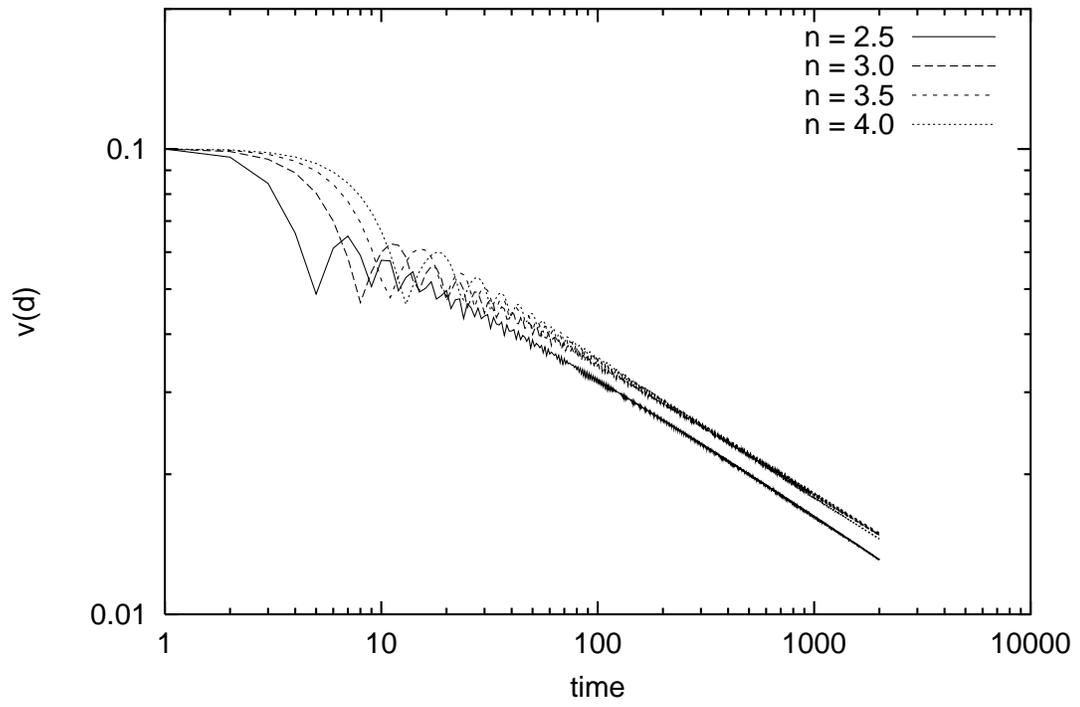
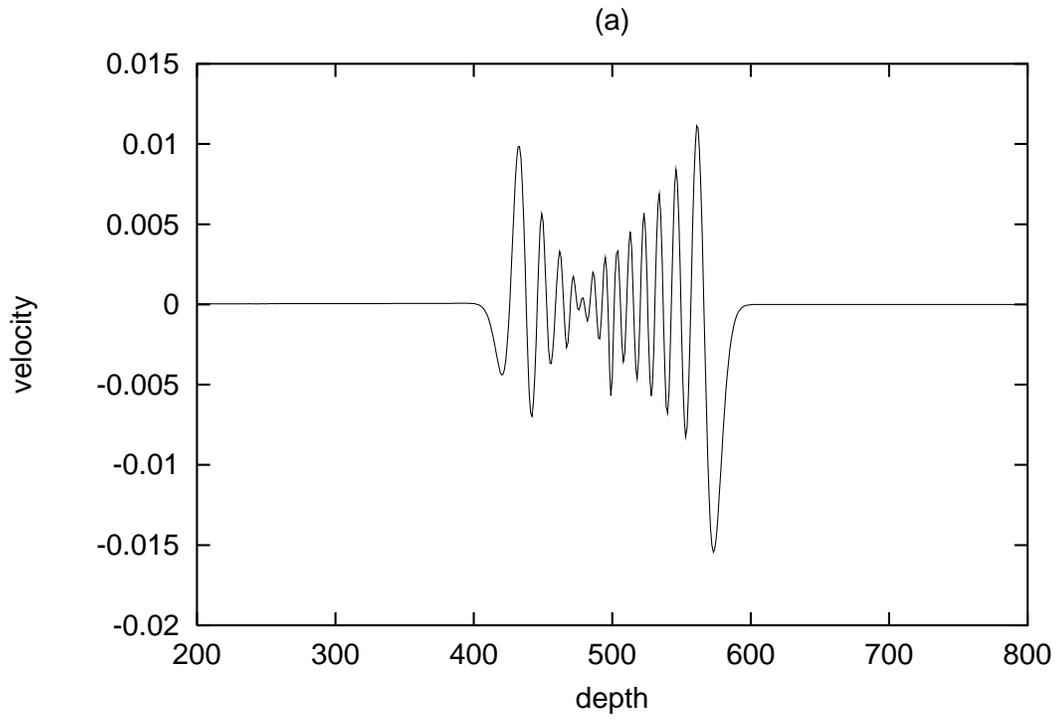
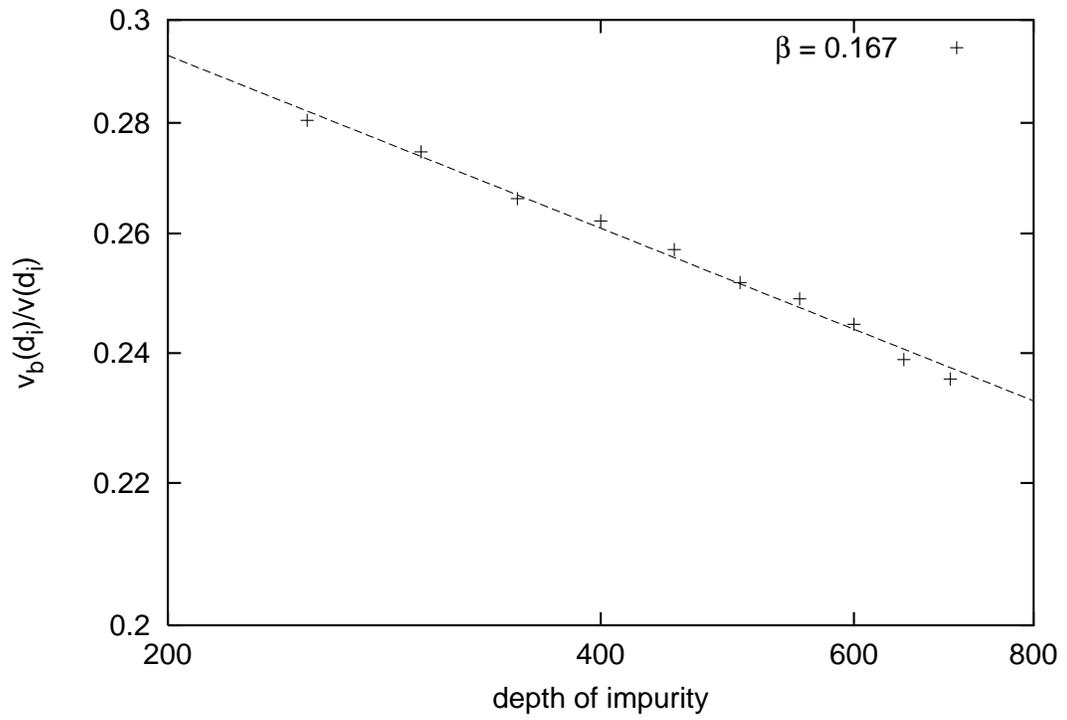
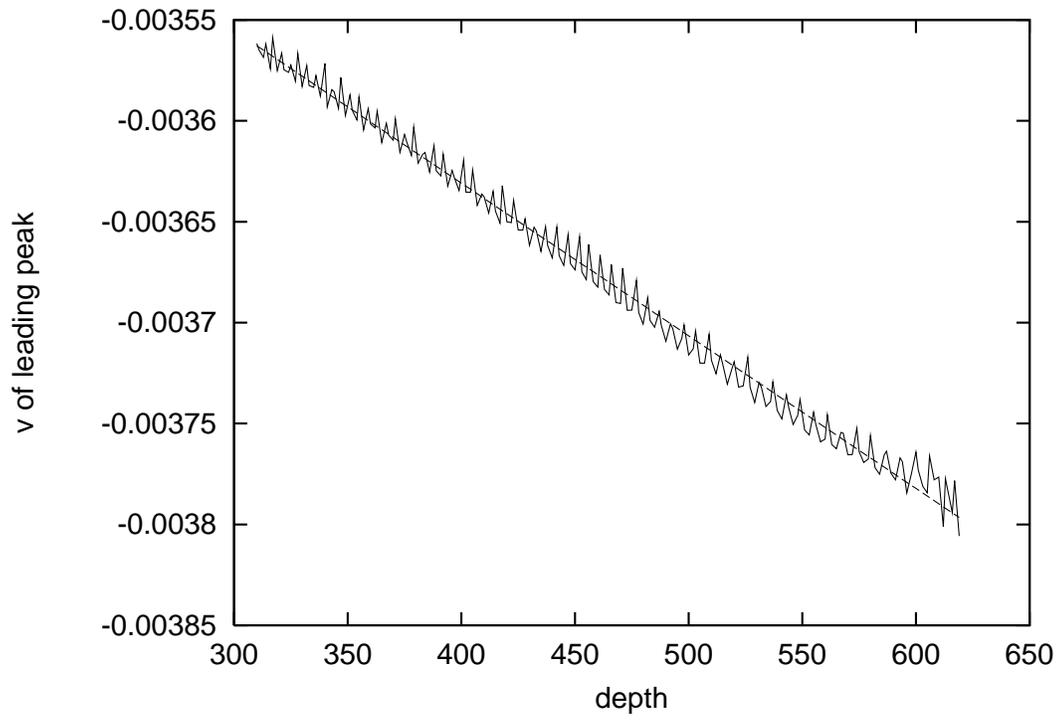
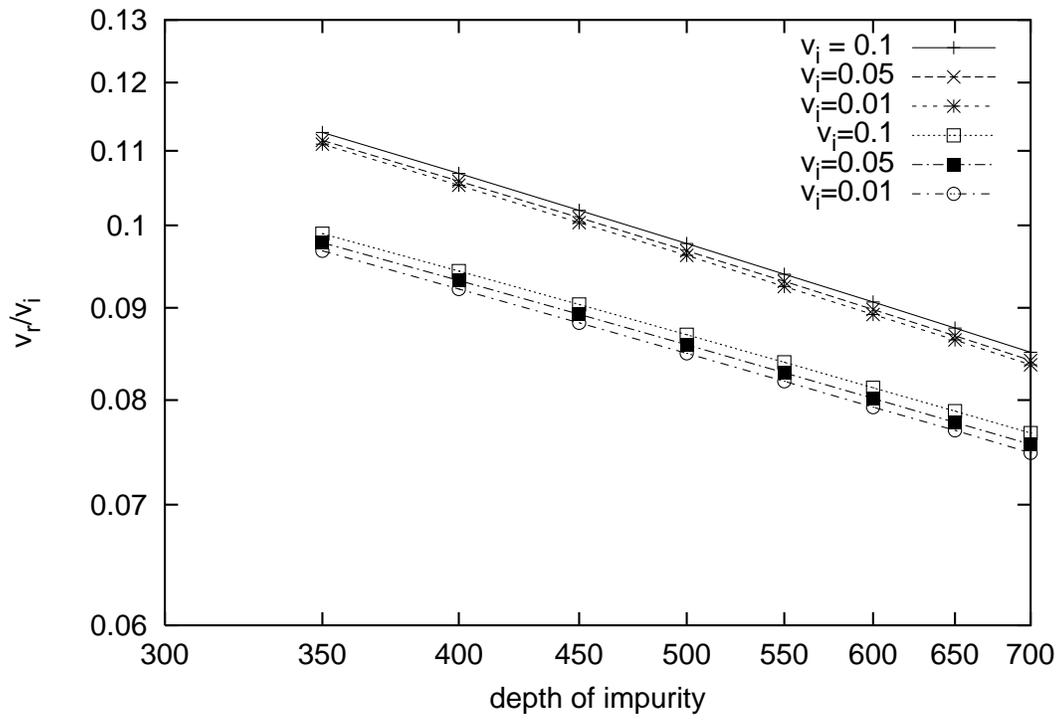
\end{document}